\documentclass[aps,prl,twocolumn,amsmath,amsfonts,amssymb,superscriptaddress]{revtex4}

\usepackage{graphicx}

\usepackage{latexsym}

 \let\b=\beta

    \let\f=\varphi

  \let\r=\rho 
\let\io=\infty

\def\to{\rightarrow}

\newcommand{\beq}{\begin{equation}}
\newcommand{\eeq}{\end{equation}}

\begin{document}

\title{A note on rattlers in 
amorphous packings of binary 
mixtures of hard spheres}

\author{I.Biazzo$^1$, F.Caltagirone$^1$, G.Parisi$^1$, F.Zamponi$^{2}$ \\
{\scriptsize $^1$Dipartimento di Fisica, Universit\`a di Roma ``Sapienza'', P.le A.Moro 5, 00182 Roma} \\
{\scriptsize $^2$CNRS LPT-ENS, 24 Rue Lhomond, 75231 Paris} \\
}

\maketitle

It has been recently pointed out in Refs.~\cite{olandesi,kyyy} that our theoretical
result for the jamming density of a binary mixture of hard spheres \cite{noi} ``apparently'' violates an upper bound that is obtained by
considering the limit where the diameter ratio $r = D_A/D_B \to \io$. We believe that this apparent contradiction is the consequence
of a misunderstanding, which we try to clarify here.

\begin{figure}[t]
\includegraphics[width=.48\textwidth]{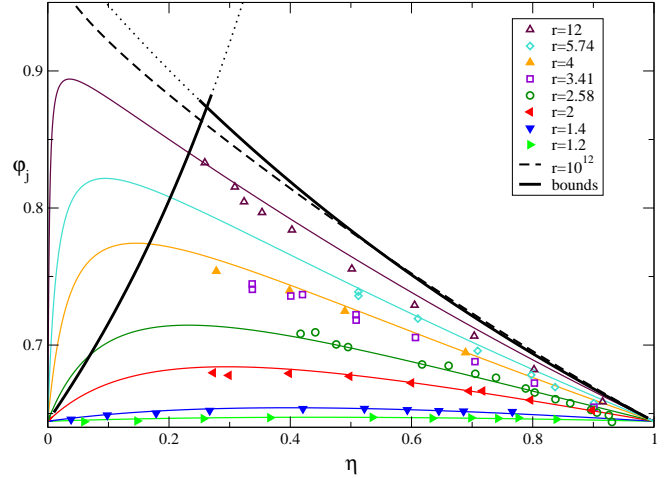}
\caption{The bound for the jamming density
in Eq.~(\ref{uno}) (thick solid lines) and the theoretical value from Ref.~\cite{noi} for $r=10^{12}$ at a complexity $\Sigma=1.7$ (dashed line).
The solid lines are predictions from our theory \cite{noi} for different values of $r$ at $\Sigma=1.7$, filled symbols are numerical results from simulations~\cite{noi} and
open symbols are experimental data from Ref.~\cite{YCW}. Note that the large $r$ - small $\eta$ region is out of the domain of validity of our theory, indeed the theoretical
curves violate the upper bound in that region.}
\label{fig:1}
\end{figure}

We will follow the notations of our paper~\cite{noi} and denote by $\eta$ the volume fraction of small particles.
Then, a very simple argument~\cite{kyyy,olandesi} leads to the following result in the limit $r\to \io$:
\beq
\label{uno}
\varphi_{j,\io}=\min\Bigl\{ \, \frac{\varphi_{j,1}}{1-\eta}\, , \, \frac{\varphi_{j,1}}{\varphi_{j,1} + \eta (1-\varphi_{j,1})}\, \Bigr\}
\eeq
where $\f_{j,1} \sim 0.64$ is the jamming packing fraction of the monodisperse system ($r=1$). 
In the following, we consider a binary mixture of hard spheres undergoing some dynamics at fixed temperature $k_B T = 1/\b$
(the precise value of which is irrelevant and fixes the time scale of the problem).
The two terms in Eq.~(\ref{uno})
correspond to very different physical situations that might happen in the limit $r\to\io$:
\begin{itemize}
\item For small enough $\eta$, the large particles form a jammed structure, while the small ones diffuse freely (as in a liquid) 
in the voids of the large ones.
\item For large enough $\eta$, all particles participate to the jammed structure, in which the large particles are ``diluted'' in a ``sea'' of small particles.
Hence, in this regime, {\it most} of the large particles only touch the small ones, while large - large contacts due to density fluctuations are rare.
It can be easily seen that, according to our theory~\cite{noi}, this qualitative picture becomes exact in the limit $r\to\io$ at fixed $\eta$ or for $\eta \to 1$ since, in these limits, 
the average coordination number of large spheres around a given large sphere at jamming, goes to zero.

\end{itemize}
These two regimes are separated by $\eta_{th} = \frac{1-\varphi_{j,1}}{2-\varphi_{j,1}} \sim 0.265$ where the two terms in Eq.~(\ref{uno}) 
are equal, see Fig.~\ref{fig:1}.

\begin{figure}[t]
\includegraphics[width=.49\textwidth]{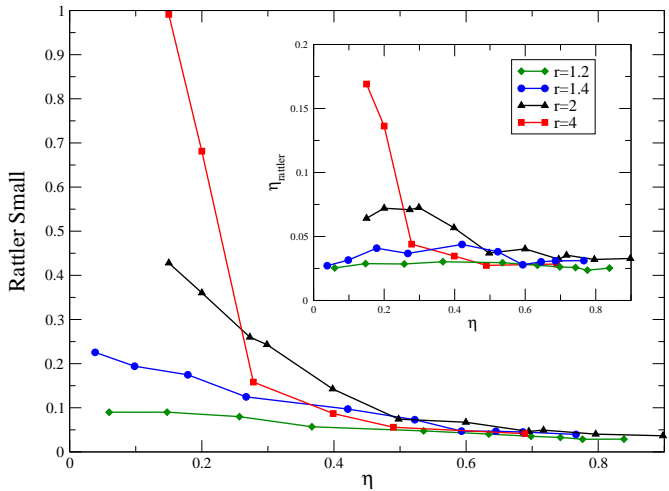}
\caption{Fraction of small rattlers $f_B = N_{rattler,B}/N_B$. 
(Inset) Volume fraction occupied by the rattlers,
defined, following~\cite{noi}, as $\eta_{rattler}=(f_A \r_A V_3(D_A) + f_B \r_B V_3(D_B))/\f$, where 
$f_\mu = N_{rattler,\mu}/N_{\mu}$ is the fraction of rattlers for species $\mu$.
}
\label{fig:2}
\end{figure}

In the {\it small cage} approximation that we use in our theory~\cite{noi,PZ10}, we assume that a particle is allowed to vibrate within a 
cage made of the particles surrounding it. Moreover we assume that the structure of the system is frozen, and the spheres 
(belonging both to the large and small components) 
can't have any kind of diffusive motion in the system. Therefore our theory cannot describe the region where the small 
particles diffuse in the voids of the large ones;
this happens for large enough $r$ and small enough $\eta$.
For this reason we avoided the large $r$-small $\eta$ region in the comparison with experimental data, 
see Fig.~\ref{fig:1} (and the caption of Fig.1 of Ref.\cite{noi}).

On the other hand, the jamming packing fraction predicted by the theory in the infinite size-ratio limit 
(shown in Fig.~\ref{fig:1} for $r=10^{12}$) is always below the upper bound given by Eq.~(\ref{uno})
in the region $\eta > \eta_{th}$ where the theory is expected to hold, see Fig.~\ref{fig:1}.

We performed simulations in order to study the 
number of rattlers in different mixtures. We used the code developed by Donev and co-workers \cite{Don1, Don2}  
based on the Lubachevsky-Stillinger algorithm \cite{Luba}. In this algorithm the spheres are compressed uniformly
by increasing their diameter, while event-driven molecular dynamics is performed at the same time. 
We carefully compress the system up to a reduced pressure $p=\beta P/ \rho = (10^{10}\div 10^{12})$ in order to easily identify the rattlers. 
The results obtained show that already for $r=4$, and concentration ratio $\eta \lesssim \eta_{th}$, the small component is not blocked,
and the fraction of rattlers among the small particles is equal to 1, see Fig.~\ref{fig:2}. 
When this does not happen, the total volume fraction of rattlers 
is always quite small and rarely exceeds $5\%$ of the total. For this reason we believe that the contribution of rattlers 
is negligible in this case.
Additionally, we want to stress that in our theory rattlers are not removed from the packing: we assume that their cage is 
vanishingly small so that they
are equivalent to the jammed particles. This is not a bad approximation since, in the region of validity of the theory, the
cages in which the rattlers are confined are very small.

\end{document}